\documentclass[twocolumn,pra,aps,showpacs,amsmath,amssymb]{revtex4}
\usepackage{graphicx}
\usepackage{bm}
\begin{document}

\title{Estimating the reduction time of quantum states}
\author{Fernando Parisio }
\affiliation{Departamento de F\'{\i}sica, Universidade Federal de Pernambuco,
Recife, Pernambuco,  50670-901, Brazil.}


\begin{abstract}
An effective description of microscopic measurements is given, in which the precise moment of probing is not determined.
Within this scenario we propose a scheme that relies on an ``attempt'' to make a forbidden simultaneous measurement 
of two incompatible observables. Although bound to failure in what concerns this goal, the process can lead to experimentally 
accessible information on a possibly non-vanishing time $\delta t$ elapsed in the collapsing of the wave function, even 
if the duration $\Delta t$ of the individual measurements is much larger than $\delta t$.

\end{abstract}
\maketitle

\section{Introduction}
\label{intro}
The relevance of the measurement problem to the general picture of quantum theory does not need emphasis. 
In particular, the collapsing versus unitary evolution of the state vector has been addressed under distinct conceptual perspectives \cite{wigner,bassi, busch, sshauer, zurek,whitaker, vera,ballentine,joos, wiseman}. Although the realization that the openness of quantum systems is probably fundamental in approaching this problem, there is no consensus about the precise role and limitations of decoherence in this regard \cite{sshauer}. In fact, on a deeper level, there is no agreement on the objective physical existence of reduction itself. A concise discussion on interpretations that either consider or disregard collapse as an ingredient of quantum theory is given in the first sections of \cite{whitaker}.
 
Regarding the collapsing scenarios, whatever the detailed mechanisms of reduction might be, if the wave function is to be  
interpreted as a meaningful physical quantity for individual particles, one has to 
acknowledge that the reduction should not be instantaneous. 
Departing from the orthodox interpretation, some models assume that the reduction of the state vector takes a non-vanishing time. Most of these works rely on stochastic mechanisms, with convenient terms added to the Schrodinger equation \cite{gisin,finite1, finite2,finite3}. The program of obtaining reduction from a dynamical equation is, to some extent, accomplished, but the time scales involved are rather arbitrary.
Alternatively, if we think of the wave function or, more generally, the density matrix as an
abstract quantity that only describes the statistical behavior of sets of particles \cite{ballentine}, then a sudden collapse is acceptable since it solely means an increase in our level of knowledge on the system, not associated to an actual instantaneous physical process. 
In hidden-variable interpretations such as that of David Bohm \cite{bohm} or in those that deny the collapse by assuming that a measurement leads to a branched many-world structure \cite{everett, griffiths}, the collapsing problem is not present. Notwithstanding, it is replaced by puzzles of equivalent magnitude.
Apart from interpretational preferences, an objective way to estimate the time scale of reduction, if it is non-zero, should be pursued. This would place the discussion on the physical reality of collapse on quantitative basis, instead of pure theoretical or philosophical ones. Also, it might enable us to rule out some of the competing conceptions on quantum measurements on more objective grounds. As pointed out by Home and Whitaker ``there should be a desire to understand more about the collapse postulate, {\it why} it often works well, its limitations, and how it should be eventually adapted or replaced." \cite{whitaker}.
In this context a natural question is: is it possible to conceive processes leading to detectable differences associated to the reduction of the wave function with different time scales? Ideally this distinction should be insensitive to the details of the dynamics of reduction and to the actual mechanisms that triggers it. 

In trying to answer this question, at least within a plausible scenario, we take into account the fact that no actual measurement device can probe a system with an infinite time accuracy and make the possibility of a finite-time reduction explicit.  
In the remainder of this paper, by a microscopic measurement we no longer mean a physically indivisible event,
but rather a composite one. 
\section{Microscopic Measurements}
\label{s2}

From the orthodox interpretation of quantum mechanics one can not scape the inference that a measurement is taken as an instantaneous event. 
In fact, two conceptually distinct abrupt processes are tacitly assumed in the measurement postulate: (i) at a precisely defined time $t_p$ the system is probed, and, at the same time, (ii) the quantum state collapses \cite{cohen}. The terms ``probing" and ``measurement" are employed in a rather interchangeable way. However, if a microscopic system is to be probed, say, by a photon emitted by an excited atom, all we know is that this should occur according to a probability distribution $p(t) \propto e^{-t/t_{decay}}$. If we wait for a time $\Delta t > t_{decay}$ the photon will be emitted  with high probability. We say that $\Delta t$ is the duration of the whole measurement process, while the probing occurs in the precise (unpredictable) instant when the system absorbs the photon. So, from an operational point of view, the measurement is associated to the time window in which the probing can potentially happen, while the probing itself is the sudden event that starts the collapse. The same kind of reasoning can be made, e. g., for conditional measurements related to tunneling \cite{uncollapsing}. We recast the measurement postulate, which is well suited for macroscopic events, as follows: 

(I) {\it Finite-time measurement and impulsive random probing:} Although we keep assumption (i), we take into account the experimental fact that any measurement has a finite duration $\Delta t$. So, the probing does occur at a precise moment $t_p$, but one can only say that this is to happen within a time window $[t_0, t_0+\Delta t]$. The time $t_p$ is taken as a random variable, obeying an appropriate distribution defined in this interval. Further, in this work, we assume that the duration of the measurement suffices to guarantee that the probing occurs. Before it, the system remains uncoupled to external degrees of freedom. 

(II) {\it Finite-time reduction:} The wave function takes a short time $\delta t$, starting from $t_p$, to be reduced. We do not make any specific statements about the time evolution during this fast reduction, except that the global state ket remains normalized. Let $| \Phi_0 \rangle \in {\cal E}_X$ be the initial state related to the relevant degrees of freedom of the probe. If  $\hat{A}$ is an observable acting in the Hilbert space of the system (${\cal E}$), with $\{| a_i \rangle \}$ being the basis of eigenstates, then any initial ket $| \psi_0 \rangle | \Phi_0 \rangle = \sum \alpha_i | a_i \rangle | \Phi_0 \rangle$, under a measurement of $\hat{A}$, is assumed to evolve in a quite general way towards one of the eigenkets, say $| a_j \rangle$, 
\begin{equation}
\label{reduction}
| \Psi(t) \rangle = \sum \alpha_i(t) | a_i \rangle | \Phi_i(t) \rangle \longrightarrow | a_j \rangle | \Phi_{f,j} \rangle \; , 
\end{equation}
where $| \Psi(t) \rangle \in {\cal E}_T={\cal E} \otimes {\cal E}_X $, $\sum |\alpha_i(t)|^2=1$, $\alpha_i(t_p)=\alpha_i$ and $\alpha_i(t_p+\delta t)=\delta_{ij}$. The above scheme is similar to the well-known von Neumann measurement, but here we have a non-instantaneous dynamics. The time dependence in $\{| \Phi_i(t) \rangle \}$ accounts for the directions taken by these normalized kets in ${\cal E}_X$ during the reduction.  We stress that no attempt is made to describe the actual processes that take place during the reduction and that the final states $\{| \Phi_{f,j} \rangle\}$ are not assumed to be orthogonal macroscopic pointers (no ``macro objectification" required \cite{bassi}). In other words, we use a generic dynamics of reduction with no assumption on the equation that leads to it. Naturally, we take the measurement to be longer than the reduction ($\Delta t > \delta t$). 

From the above it is clear that one is not trying to provide a solution to the measurement problem. Rather, as will become evident, we seek physical predictions that may enable us to rule out either instant collapse or finite-time reduction. If the latter is ultimately verified it is also important to estimate its time scale.

We proceed by considering two incompatible observables $\hat{A}$ and $\hat{B}$, for which $\hat{A}| a_i \rangle=a_i|a_i \rangle$ and $\hat{B}| b_i \rangle=b_i|b_i \rangle$. The probability of obtaining an arbitrary pair $a_i$ and $b_j$ as a result of successive projective measurements of $\hat{A}$ and $\hat{B}$ is different from that of getting $b_j$ and $a_i$ for measurements of $\hat{B}$ and $\hat{A}$, i.e., the conditional probabilities $P(a_i|b_j)$ and $P(b_j|a_i)$ do not coincide, in general (this, of course, is not restricted to quantum mechanics). Most importantly, $P(a_i|\hat{B})$, the probability of obtaining $a_i$ knowing that $\hat{B}$ was measured and $P(a_i)$ do not coincide. Thus, a simultaneous measurement of the associated physical quantities is ill defined in the strict orthodox framework. In spite of this fact, several weaker definitions of joint measurements of incompatible observables can be found in the literature \cite{busch2,botero,yuen,ashhab,wei}. Let us analyze in detail what are the possible consequences of {\it trying} to do such a simultaneous measurement assuming the validity of statements (I) and (II). 

An important proviso arises when one looks at the classical experiment by Stern and Gerlach. Trying to observe $\hat{S}_z$ and $\hat{S}_x$ simultaneously amounts to superimpose, during a certain time interval, two macroscopic magnetic fields in the $z$ and $x$ directions. This naive approach results in an overall field, whose direction $u$ is intermediate between those of the $x$-axis and $z$-axis, and we simply execute a single measurement of $\hat{S}_u$. We assume that, no matter the nature of the probes, they act independently \cite{gisin} interacting with the system at the quantum level. In the present case, this can be done only because these measurements are assumed to be non-selective, without any macroscopic readout.

\section{Two-level systems}
\label{s3}   

In this section, to focus only on the implications of our hypothesis, we address a simple two-level system. Any observable can be thought as a combination of the identity operator and the pseudo-spin Pauli 
operators ($\hat{\sigma}_1$, $\hat{\sigma}_2$, $\hat{\sigma}_3$). Let $\{| +\rangle, | - \rangle \}$ be the orthonormal basis of eigenstates of $\hat{\sigma}_3$, as usual. Although this is not essential to validate our conclusions, due to the delicate nature of these microscopic measurements it is natural to conceive them as a series of individual realizations (so that no coherences are expected in the final statistical operator).  
We choose the two incompatible observables to be measured simultaneously as $\hat{\sigma}_1$ and $\hat{\sigma}_3$, and take the initial individual states as $| + \rangle$, leading to the density operator
\begin{equation}
\nonumber
\hat{\rho}_0=| + \rangle \langle +| \;. 
\end{equation}
To simplify the presentation we assume that both probings have uniform probability densities ${\Delta t}^{-1}$ over $[t_0, t_0+\Delta t]$.
Forcing two devices to make a simultaneous measurement is, in fact, to adjust them to start the process at, say, $t_0$ and finish it at $t_0+\Delta t$, so that both occur within $[t_0, t_0+\Delta t]$. According to our model, however, the probing associated to $\hat{\sigma}_1$ happens at a precise time $t_p^{(1)}$, while that of $\hat{\sigma}_3$ occurs at $t_p^{(3)}$.
The assumption that the devices make measurements with the same duration $\Delta t$ is by no means essential and, again, aims to simplify the presentation. Extensions to situations with $\Delta t^{(3)}\ne \Delta t^{(1)}$ are straightforward. 

\begin{figure}[ht]
\includegraphics[width=6cm]{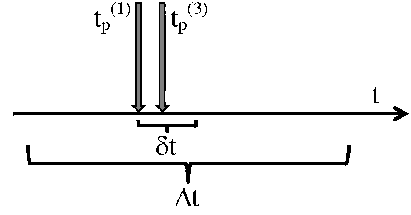}
\caption{Sequence of nearly coincident probings of the type $\hat{\sigma}_1$-$\hat{\sigma}_3$.}
\label{fig1}
\end{figure}

Let us initially assume that $\delta t=0$, i. e., the reduction is, in fact, instantaneous. 
Since the probability distributions for the probings are equal, we have $t_p^{(1)}<t_p^{(3)}$ in $50 \%$ of the realizations (asymptotically), in which case the state after the first probing is either $(|+ \rangle + | - \rangle)/\sqrt{2}$ with probability 0.5, or $(|+ \rangle - | - \rangle)/\sqrt{2}$ with the same probability. The second non-selective individual measurements lead to a final density operator given by $\hat{\rho}=0.5| + \rangle \langle +| + 0.5| - \rangle \langle -|$. In the remaining events we have $t_p^{(3)}<t_p^{(1)}$, so that the first probing does not change the initial state, the later one, of $\hat{\sigma}_1$, leading equally to $\hat{\rho}=0.25(| + \rangle + | -\rangle)( \langle +| +\langle -|) + 0.25(| + \rangle -| - \rangle)( \langle +|- \langle -|)=0.5| + \rangle \langle +| + 0.5| - \rangle \langle -|$. Thus, from an ensemble point of view, the result is insensitive to the ordering of the probings and the final reduced density operator is simply that of an unpolarized statistical mixture:
\begin{equation}
\label{collapse}
\hat{\rho}_{coll}=\frac{1}{2}| + \rangle \langle +| + \frac{1}{2}| - \rangle \langle -| \;.
\end{equation} 
Now we fully take hypothesis (II) into account, by setting $\delta t>0$. In this case the analysis is a bit more subtle.
It is still true that occurrences with $t_p^{(1)}<t_p^{(3)}$ and those with $t_p^{(3)}<t_p^{(1)}$ are equally likely, 
but the time separation between the two probings is now relevant. Thus, we define the random variable $y=|t_p^{(1)}-t_p^{(3)}|$, which, in turn, is not uniform on $[0, \Delta t]$. The probability distribution $p'(y)$
is given by
\begin{eqnarray}
\nonumber
p'(y)=\int_{\square} {\rm d}^2{t'}_p p[{t'}_p^{(1)}] \, p[{t'}_p^{(3)}] \delta [y-|{t'}_p^{(1)}-{t'}_p^{(3)}|]\\
\label{py}
=\frac{2}{\Delta t^2} (\Delta t-y) \;,
\end{eqnarray}
where the integration is on the square with side length $\Delta t$ in the plane ${t'}_p^{(1)}$ - ${t'}_p^{(3)}$, with $p[{t'}_p^{(1)}]=p[{t'}_p^{(3)}]={\Delta t}^{-1}$.
Therefore, the probability that the second probing occurs {\it during the reduction} started by the first 
one is
\begin{equation}
P(y <\delta t)=2 \left( \frac{\delta t}{\Delta t}\right)-\left( \frac{\delta t}{\Delta t}\right)^2 \;.
\end{equation}
The key point is that, although we do not have any control of the time elapsed between the two impulsive probings, these nearly coincident events will occur with a certain frequency due to statistical robustness, for a sufficiently large number of repetitions. For realizations in which the time separation between the probings oversteps $\delta t$ the result is exactly that given by (\ref{collapse}).
Therefore, the total density operator associated to a fast reduction scenario should have the general form $\hat{\rho}_{red}^{T}=[1-P(y <\delta t)]\hat{\rho}_{coll}^{T} +P(y <\delta t)\tilde{\rho}^{T}$, were $\hat{\rho}_{coll}={\rm Tr}_X(\hat{\rho}_{coll}^{T})$
and $\tilde{\rho}^{T}$ is to be determined. We write $\tilde{\rho}^{T}=0.5\tilde{\rho}^{T}_{1-3}+0.5\tilde{\rho}^{T}_{3-1}$, to distinguish the contributions from the two possible sequences of probings. Since $\hat{\rho}_0=| + \rangle \langle + |$, if the sequence is $\hat{\sigma}_3$-$\hat{\sigma}_1$ there is no initial reduction and it is easy to show that we get nothing but ${\rm Tr}_X(\tilde{\rho}^{T}_{3-1})=\hat{\rho}_{coll}$. 
Thus,
\begin{equation}
\label{r1b}
\hat{\rho}_{red}=\left[1-\frac{1}{2}P(y <\delta t)\right]\hat{\rho}_{coll} +\frac{1}{2}P(y <\delta t){\rm Tr}_X(\tilde{\rho}^{T}_{1-3}) \;.
\end{equation} 
We now address the non-trivial sequence in which the probing related to $\hat{\sigma}_1$ happens first, see fig. \ref{fig1}.
During the short time elapsed before the probing corresponding to $\hat{\sigma}_3$ happens, the global state is either $| \Psi(t) \rangle$ or $| \Psi'(t) \rangle$, where
\begin{eqnarray}
\label{states}
\nonumber
| \Psi(t) \rangle=a_{+}(t)| + \rangle | \Phi_+(t) \rangle + a_{-}(t)| - \rangle | \Phi_-(t) \rangle \;,\\
| \Psi'(t) \rangle=b_{+}(t)| + \rangle | \Phi'_+(t) \rangle- b_{-}(t)| - \rangle | \Phi'_-(t) \rangle  \;,
\end{eqnarray}
with equal probabilities. In the first case the state at $t=t_p^{(1)}+\delta t$ would be $(| + \rangle + |- \rangle)| \Phi_f \rangle/\sqrt{2}$,
while in the second one $(| + \rangle - |- \rangle)| \Phi'_f \rangle/\sqrt{2}$.
Accordingly, we have
$a_{+}(t_p^{(1)})=1 \; , \;  a_{-}(t_p^{(1)})=0 $,
with the reduction dynamics being disturbed by the second (incompatible) probing before the conditions
$| \Phi_+(t_p^{(1)}+\delta t) \rangle=| \Phi_-(t_p^{(1)}+\delta t)\rangle=| \Phi_f \rangle$,
$a_{+}(t_p^{(1)}+\delta t)=1/\sqrt{2}$ , $a_{-}(t_p^{(1)}+\delta t)=1/\sqrt{2}$,
are attained, since $t_p^{(1)}<t_p^{(3)}<t_p^{(1)}+\delta t$ (a set of analogous conditions being valid for $b_{\pm}$ and $| \Phi'_{\pm} \rangle$).
Since the intermediate states are equally likely the probability density of getting $| + \rangle$ after a measurement of $\hat{\sigma}_3$ is $0.5(|a_+(t)|^2+|b_+(t)|^2)$, while for $| - \rangle $, it is proportional to $0.5(|a_-(t)|^2+|b_-(t)|^2)$. In order to get the statistical operator, we have to integrate over all possibilities with the appropriate weight. 
The probability that the second probing occurs between $t$ and $t+{\rm d}t$ with $t_p^{(1)}<t<t_p^{(1)}+ \delta t$ is $ p'(y){\rm d}y/P(y <\delta t) $. Therefore, the density operator $\tilde{\rho}^{T}_{1-3}$ describing the two subensembles that originate from (\ref{states}) reads \cite{comment}:
\begin{equation}
\tilde{\rho}^{T}_{1-3}=\int_0^{\delta t}\frac{p'(t) {\rm d}t}{P(t <\delta t)} \tilde{\Lambda}\;,
\end{equation}
with the first probing set at $t_p^{(1)}=0$ and 
\begin{eqnarray}
\nonumber
\tilde{\Lambda}=\frac{1}{2}| + \rangle (|a_{+}|^2|\Phi_{+}\rangle \langle  \Phi_{+}|  +|b_{+}|^2|\Phi'_{+}\rangle \langle \Phi'_{+}|)\langle +|\\ 
+\frac{1}{2}| - \rangle (|a_{-}|^2|\Phi_{-}\rangle \langle  \Phi_{-}|  +|b_{-}|^2|\Phi'_{-}\rangle \langle \Phi'_{-}|)\langle -|\;,
\end{eqnarray}
where the dependence on $t$ was omitted.
By tracing over all degrees of freedom not belonging to ${\cal E}$ we get
\begin{eqnarray}
\nonumber
\tilde{\rho}_{1-3} = \gamma|+ \rangle \langle +|+(1-\gamma)|- \rangle \langle -| \;,
\end{eqnarray} 
where 
\begin{equation}
\gamma = \left( 1-\frac{\tau}{2} \right)^{-1} \int_{0}^{1}{\rm d}x(1-\tau x)|c_{+}(x)|^2 \;,
\end{equation}
with $|c_+(t/\delta t)|^2 \equiv 0.5(|a_+(t)|^2+|b_+(t)|^2)$. We used the assumption that the state ket remains normalized ($|a_-(t)|^2=1-|a_+(t)|^2$, $|b_-(t)|^2=1-|b_+(t)|^2$)
and defined $\tau=\delta t/ \Delta t$. Note that all the influence of the specific time
evolution of the system during the reduction is contained in the
single number $\gamma$, which is weakly dependent on the particular path followed on the Bloch sphere. Since the maximum value $|c_{+}(x)|^2$ can reach is $1$, it is straightforward to 
show that $0\le \gamma \le 1$, no matter how intricate the actual reduction dynamics may be. 
Furthermore, instantaneous collapse can be introduced not only by directly setting
$\delta t =0$ but also by assuming that $|c_{+}(x)|$ goes suddenly to $1/\sqrt{2}$ after $t_p^{(1)}$. This leads to $\gamma=1/2$ which, 
consistently, yields $\tilde{\rho}=\hat{\rho}_{coll}$. 
In general we obtain 
$\hat{\rho}_{red}=(1-\tau+\tau^2/2)\hat{\rho}_{coll} +( \tau-\tau^2/2)[\gamma|+\rangle \langle +|+(1-\gamma)|- \rangle \langle -|]$.
The final result is then
\begin{eqnarray}
\label{r5}
\nonumber
\hat{\rho}_{red}=\frac{1}{2}\left[ 1+\left( \tau-\frac{\tau^2}{2}\right)(2\gamma-1)\right]|+ \rangle \langle +|\\
+ \frac{1}{2}\left[ 1-\left( \tau-\frac{\tau^2}{2}\right)(2\gamma-1)\right]|- \rangle \langle -| \;,
\end{eqnarray}
that describes a completely unpolarized ensemble with respect to $\hat{\sigma}_1$ and $\hat{\sigma}_2$ \cite{comment}, but with slightly distinct probabilities for the outcomes $| + \rangle$ and $| - \rangle$ in a measurement of $\hat{\sigma}_3$, which could, in principle, be experimentally detected. For $0\le \gamma < 1/2$ we have $N_->N_+$ and for $1/2 < \gamma \le 1$, $N_-<N_+$, where $N_{\pm}$ is the counting associated to $| \pm \rangle$. Of course these conditions would be inverted had we chosen $| - \rangle$ as the initial state. We stress that only this later measurement must be macroscopic, with the results related to pointers. 
The associated probabilities are
\begin{equation}
\nonumber
{\cal P}_3(+)={\rm Tr}(\hat{\rho}_{red} |+ \rangle \langle +|)=\frac{1}{2}+ \frac{1}{2} \left( \tau-\frac{\tau^2}{2}\right)(2\gamma-1)\;,
\end{equation}
and ${\cal P}_3(-)=1-{\cal P}_3(+)$. So that, for a large number $N$ of realizations the difference between results $| + \rangle$ and $| - \rangle$ is given by
\begin{equation}
\label{N}
\Delta N =N_+ - N_-=\left( \tau-\frac{\tau^2}{2}\right)(2\gamma-1)N\;.
\end{equation}
In the limit of very fast reduction we have $\delta t << \Delta t$, i. e., $\tau << 1$, and the above quantity becomes $\Delta N =\tau(2\gamma-1)N$ with $\gamma$ being simply $\gamma = \int_{0}^{1}|c_{+}(x)|^2 {\rm d}x$, which is independent of $\tau$.   
\section{Statistical fluctuations}
Since one can not demand $N \rightarrow \infty$ in actual experiments, statistical fluctuations will show up, and we have a finite probability to observe differences between $N_+$ and $N_-$ even in a scenario of instantaneous collapse. 
If we denote these statistical fluctuations by $\Delta {\cal N}$,
for $\delta t \ne 0$ we should be able to generate results with $\Delta N>>\Delta {\cal N}$ in order to obtain a clear evidence
of finite-time reduction. This is indeed possible since $\Delta {\cal N} \sim \sqrt{N}$ and $\Delta N \sim N$, therefore, reliable data could be obtained for 
\begin{equation}
\label{condition}
N>> \left[\frac{\Delta t}{\delta t (2\gamma -1)}\right]^2\;.
\end{equation}
If $\Delta N \sim \Delta {\cal N}$ for arbitrary $N$, then the finite-time reduction, as presented here, can be ruled out.
In practice, if $\Delta N \sim \Delta {\cal N}$ for a large $N$ then $\tau_{max} \sim N^{-1/2}$ gives an upper bound
for the reduction time. 

To illustrate the application of the general ideas just presented, suppose one intends to investigate $\delta t$ in the range of, say, tenths of picoseconds with the time resolution $\Delta t$ of the devices being about 1 ns ($\tau \sim 0.01$). The particular form of $c_{+}(t/\delta t)$ is not of critical importance and we take an exponentially decaying function satisfying the appropriate boundary conditions:
\begin{equation}
\nonumber
c_{+}(t/\delta t)=\frac{(2-\sqrt{2})e^{-\frac{t}{\delta t}+\frac{t_p^{(1)}}{\delta t}} +\sqrt{2}-2e^{-1}}{2(1-e^{-1})}\;,
\end{equation}
for which $\gamma \approx 0.695$ and condition (\ref{condition}) reads $N>>65.000$. We remark that for a drastically distinct behavior, a linearly decaying function, we find $\gamma \approx 0.735$, that would not change the order of magnitude of the results. Let us take $N=800.000$, which, of course, poses a practical difficulty. If $\delta t$ is indeed of order of $10$ ps, one should get $\Delta N \approx 3100$, while $\Delta {\cal N} \approx 900 $. Considering the worst scenario were statistical fluctuations tend to compensate for $\Delta N$, we would have a difference of $2200$. The chance that such a difference is caused by pure statistical fluctuations is about $0.05\% $ , constituting a strong evidence for finite-time reduction in the scale of tenths of picoseconds. If we find $\Delta N \approx \Delta {\cal N}$, then $0 \le \delta t < 10$ ps. 

\section{Conclusion}
\label{s4} 
The main point of this work is to propose that, in addition to the search for the dynamical mechanisms of collapse, we may look for schemes capable of delivering information on the time scale of reduction, no matter its details. For both, those that find more natural to face quantum mechanics via interpretations that do not need the concept of physical collapse and those that think of the state vector as an actual entity for individual systems, this kind of approach may open an opportunity to more objective discussions.
Although the unavoidable experimental imperfections were disregarded here,
the previous results show that these schemes can be realized within the proposed scenario. The assumption of perfect measuring devices can be partially compensated by the fact that our choice of the probing probability distribution $p(t)={\Delta t}^{-1}$, is not a favorable one for the occurrence of two probings within a given time window. This simplifying hypothesis, arbitrary as it is, aimed to facilitate the proof-of-principle that this work intended to give. More realistic distributions, e. g., associated to spontaneous decay of excited atoms or to tunneling, would possibly lead to a clearer distinction between $\Delta N$ and $\Delta {\cal N}$. This point along with a more detailed analysis for systems with larger Hilbert spaces will be investigated shortly.
As a final comment, we recall that our reasoning is strictly limited to the non-relativistic realm. 
Since the two probings do not have any causal relation, there are inertial frames in which the ordering is swapped, e. g., $\hat{\sigma}_1$-$\hat{\sigma}_3$ to $\hat{\sigma}_3$-$\hat{\sigma}_1$, leading to non-covariant results.

\begin{acknowledgments}
The author thanks F. Brito and A. M. S. Mac\^edo for stimulating discussions on this work. Financial support 
from the Brazilian agencies CNPq and FACEPE (APQ-1415-1.05/10) is acknowledged.
\end{acknowledgments}


\begin{thebibliography}{99}
\bibitem{comment} This is a consequence of the individual character of each experiment. 
Had we considered a beam containing a statistical mixture of the states in (\ref{states}), 
$\tilde{\rho}^{T}_{1-3}=\frac{1}{2}\int_0^{\delta t}\frac{p'(t){\rm d}t}{P(t <\delta t)}(|\Psi(t) \rangle 
\langle \Psi(t)|+|\Psi'(t) \rangle \langle \Psi'(t)|)$, 
coherences would appear and $\lambda|+ \rangle \langle -| + \lambda^*|- \rangle \langle +|$
should be added to Eq. (\ref{r5}), with $\lambda=\frac{1}{2}\int_0^{\delta t}\frac{p'(t){\rm d}t}{P(t <\delta t)}
[a_+(t)a_-^*(t)\langle \Phi_-(t)| \Phi_+(t)\rangle
+b_+(t)b_-^*(t)\langle \Phi'_-(t)| \Phi'_+(t)\rangle]$. This, however, would not cause any difference in the result 
of the last measurement of
$\hat{\sigma}_3$.
\end{thebibliography}
\end{document}